\documentclass[a4paper0,12pt]{article}

\usepackage{latexsym,amsfonts,amsthm,amsmath,amscd,amssymb,color}
\usepackage{graphicx}
\usepackage{amsbsy}
\usepackage{cite}
\usepackage[all]{xy}

\begin{document}

\newcommand{\sect}[1]{\section{#1}}
\newcommand{\subsect}[1]{\subsection{#1}}
\newcommand{\subsubsect}[1]{\subsubsection{#1}}
\renewcommand{\theequation}{\thesection.\arabic{equation}}
\setcounter{equation}{0}
\numberwithin{equation}{section}

\newtheorem{lemma}{Lemma}[section]
\newtheorem{thm}[lemma]{Theorem}
\newtheorem{rem}[lemma]{Remark}
\newtheorem{rems}[lemma]{Remarks}
\newtheorem{prop}[lemma]{Proposition}
\newtheorem{cor}[lemma]{Corollary}
\newtheorem{conj}[lemma]{Conjecture}
\newtheorem{ex}[lemma]{Exercise}
\newtheorem{example}[lemma]{Example}
\newtheorem{defn}[lemma]{Definition}
\newtheorem{fact}[lemma]{Fact}
\newtheorem{ques}[lemma]{Question}
\newtheorem{hyp}[lemma]{Hypothesis}

\setlength{\unitlength}{1cm}
\newcommand{\be}{\begin{eqnarray}}
\newcommand{\ee}{\end{eqnarray}}
\newcommand{\bee}{\begin{eqnarray*}}
\newcommand{\eee}{\end{eqnarray*}}
\newcommand{\pmat} {\begin{array}}
\newcommand{\ep} {\end{array}}
\newcommand{\ra}{\rightarrow}
\newcommand{\sse}{\subsection}
\newcommand{\+} {\`}
\newcommand{\es} {\epsilon}
\newcommand{\pa} {\partial}
\newcommand{\ddh} {\dd h}
\newcommand{\al} {\alpha}
\newcommand{\dd} {\dot}
\newcommand{\hb} {\hbar}
\newcommand{\asy}{{\cal O}

\newcommand{\ind}{\hskip 0.5cm}}

\def\spazio#1{\vrule height#1em width0em depth#1em}

\title{Bender-Wu singularities}
\author{\textbf{Riccardo Giachetti}\\Dipartimento di Fisica e Astronomia, Universit\+a di 
	Firenze,\\ 50019 Sesto Fiorentino.  I.N.F.N. Sezione di Firenze. \\\\ \textbf{Vincenzo 
		Grecchi}\\Dipartimento di Matematica, Universit\+a di Bologna, 40126 Bologna. \\GNFM, 
	INdAM, Roma}
\maketitle
\textbf{Abstract}\\
\small{We consider a  
	family of quantum  Hamiltonians $H_\hb=-\hb^2\,(d^2\!/dx^2) +V(x)$, $x\in\mathbb{R},$ $\hb>0,$ 
	where 
	$V(x)=i(x^3-x)$ 
	is an imaginary  double well potential. We prove the existence of infinite eigenvalue  
	crossings with the selection rules
	of the eigenvalue pairs taking part in  a  crossing.   
	This is a semiclassical localization effect. 
	The eigenvalues at the crossings accumulate at a critical energy for some of the 
	Stokes lines.}
\section{Introduction and statement of the results}

In this paper we consider some spectral properties of the cubic oscillator described by the family 
of closed Hamiltonians
\be H_\hb=-\hb^2\,(d^2\!/dx^2)+V(x)\,,\,\,x\in\mathbb{R}\,,~~\,\mathrm{with}\,~\, V(x)=i(x^3- 
x)\,~ 
\mathrm{and} 
~\,\hb>0  
\label{RKHKC1}\ee 
defined on the domain $\mathcal{D}_H=D(\,d^2/dx^2)\bigcap D(|x|^3)$. More precisely we prove the 
existence
of an infinite number of crossings of its eigenvalues $E_n(\hbar)$ (or \emph{levels}) and we 
specify the 
selection rules for the two level pairs involved in a given crossing. 
Actually, the crossing is possible 
only for a pair of levels because for large $\hb$ only one of the nodes is unstable and the 
localization can be in one of only two wells.  Since the spectrum is simple, the crossing of two 
levels implies the non holomorphy  of the
functions $E_n(\hb)$ involved. In particular, at the crossing there is a  branch 
point called Bender-Wu singularity \cite{BW,BG,A}.
For real $\hbar>0$ the
Hamiltonians (\ref{RKHKC1}), and in particular the potential, are $PT$-symmetric 
operators \cite{BOG,BB,SH,GGD}. As  $V(x)$ has two stationary points at  $x_\pm=\pm 1/\sqrt{3}$ 
we will speak of a
$PT$-symmetric double well oscillator, with states that possibly localize at one of the wells or at 
both of them. 

Anharmonic oscillators are basic non solvable models in quantum mechanics. They pose a 
summability 
problem analogous to the one encountered in quantum field theory and therefore they have been 
extensively investigated since a long time \cite{LM,SI,GGS,SI1,C}. The cubic oscillator also has 
been 
studied by several 
approaches. We use here the nodal analysis \cite{SHA,GM}, 
namely the study of the confinement of zeros of
the entire eigenfunctions $\psi_n(x)=\psi_n(x,\hb)$ (or\textit{ states}) and of their derivatives. 
 For that, we use   the Loeffel-Martin method for the 
 control of the zeros  \cite{LM}, the asymptotic behavior of Sibuya \cite{S}, the 
 semiclassical accumulation of the zeros in some of the Stokes lines \cite{GI} and the exact 
 semiclassical quantization. Moreover the results of perturbation theory \cite{GM,C} 
 and the unique summability of the perturbation series are also relevant.
In such a way we believe 
that we can draw an exhaustive picture of the level crossing. We must however mention that the
semiclassical theory \cite{BG,BW,A} has given good results for low values of the parameter 
$\hbar$ up to the crossing value, slightly  extended by the exact semiclassical theory 
\cite{V,DDP,DT} to larger, although not very large values of  $\hbar$. A still different 
rigorous technique is  found 
in \cite{E,EG}. Of course all these treatments are very useful and complementary to ours, which 
was 
presented in \cite{GG} in a preliminary and not completely rigorous form in some parts.

We recall that analytic families of self-adjoint Hamiltonians     
\cite{K} cannot present level crossings with Bender-Wu singularities. This is also the case of many families of
$PT$-symmetric Hamiltonians with a single well potential \cite{BG,BOG,BB}.
In the following we also will take advantage of the results established  for two more 
families of Hamiltonians, defined on the same domain $\mathcal{D}_H$ as (\ref{RKHKC1}), with cubic 
potentials different from $V(x)$ but related to $H_\hbar$ by  changes of parametrization. 
The first one is an analytic family of type A \cite{K} single well complex cubic oscillators
\be \widetilde{H}_\beta=-\,(d^2\!/dx^2)+x^2+i\sqrt{\beta}x^3,\,\quad\beta\neq 0, 
\,\quad |\arg(\beta)|<\pi\, \label{Hbeta}\ee 
which will be used to identify a semiclassical level as a continuation of a perturbative one.
All the levels  $\widetilde{E}_n(\beta)$ of $\widetilde{H}_\beta$ and the corresponding states 
$\widetilde{\psi}_n(\beta)$
are perturbative with labels $n$ determined by the number of zeros which are stable
at $\beta=0$ (or \emph{nodes}). The $n+1$ stationary points of $\widetilde{\psi}_n(\beta)$
stable at $\beta=0$ will be called \emph{antinodes}. 
Notice that
$H_{\beta=0}$ reduces to an harmonic oscillator whose states are concentrated in the interval
$[\, -\sqrt{E},\sqrt{E}\,]$, namely about their antinodes.
In the paper \cite{GM} A. Martinez and one of 
us (V.G.) have extended to (\ref{Hbeta}) the proof of the absence of crossings. 
For later use we show here the relationship between (\ref{RKHKC1}) and (\ref{Hbeta}). We first
make a translation of $H_\hbar$ centered at each of the two wells letting 
$x=y+x_\pm=y\pm1/\sqrt{3}$ and defining the two isospectral Hamiltonians
 \be {H}_\hb^\pm=-\hb^2\,(d^2\!/dy^2)+i(y^3\pm\sqrt{3} y^2)\pm {E_0},\,\qquad {E_0}= -ic\,,\quad c= 
 \frac{2}{3\sqrt{3}} \label{Hpm} \ee
In order to use the perturbation theory  \cite{GM} we make the dilations \cite{Combes}:
$\,y=\lambda^\pm(\hb) z$ with $\lambda^\pm(\hb)=3^{-1/8}\,\exp(\mp i\pi/8)\,\sqrt \hb\,.$
Letting
\be c^\pm=3^{1/4}\sqrt{\pm i},\,\,\qquad  \beta^\pm(\hb)= 3^{-5/4}\,\exp(\mp\, i5\pi/4)\,\hb 
\label{betapm}\ee
we find  isospectrality  ($\sim$) between $\widetilde{H}_\beta$ and $H_\hbar$ in the form 
\be\widetilde{H}_{\beta^\pm(\hb)} \sim (1/\hb c^\pm)\, H_\hb^\pm\,\mp E_0\,, \qquad
\widetilde{E}_n(\beta^\pm(\hb)) = (1/\hb c^\pm)\, E_n^\pm(\hb)\,\mp E_0
\label{isospettrobeta}\ee
It can be proved rather easily that for $\hbar$ small enough the levels $ E_n^\pm(\hb)$
are respectively obtained from $E_n(\hb\exp(\pm i\pi/4))$ by analytic continuations
along paths in the complex plane of $\hb$ with $|\hb|$ fixed \cite{C}. For small  $\hbar$
we also have that  $E_n^\pm(\hb)$ are non-real and complex conjugated. 

The necessity of level crossings is determined by comparing the behavior of the levels
$E_n(\hbar)$ for large values of $\hbar$ with the behavior of $E^\pm_n(\hbar)$ for small $\hbar$.
It is very useful to introduce  a parametrization of the Hamiltonians (\ref{RKHKC1})  more suited 
when $\hbar$ becomes large. Again on the domain $\mathcal{D}_H$ we thus define the family
\be \widehat{H}_\al=-\,(d^2\!/dx^2)+i(x^3+\al x),\qquad\al\in\mathbb{C}\label{KHH}\ee
with  levels $\widehat{E}_n(\al)$ and states  $\widehat{\psi}_n(\al)$. 
The simple regular dilation 
\be x\ra \lambda x\,,\qquad \lambda=\sqrt{-\al} =\hb^{-2/5}\label{dilalambda}\ee 
gives the relation
\be \widehat{E}_n(\al(\hb))= \hb^{-6/5}\,E_n(\hb),\qquad \al=-\hb^{-4/5}\leq 
0\,\,,\label{RHT_new}\ee
and, in particular, $ \hb^{-6/5}E_n(\hb)\ra\widehat{E}_n(0)$ as $\hb\ra+\infty$. 
The eigenvalues $\widehat{E}_n(\al)$ are thus holomorphic in a neighborhood of the origin and 
on the sector
\be\mathbb{C}_\al=\{\al\in\mathbb{C},\, \al\neq 0, |\arg(\al)|< 4\pi/5\}.\label{ST}\ee

In view of (\ref{isospettrobeta}) we have isospectrality between 
$\widetilde{H}_\beta$ and $\widehat{H}_\al$. Indeed the relations 
\be\widetilde{H}_\beta\sim 
\beta^{1/5}\,\widehat{H}_{\al(\beta)}-\frac{2}{27\beta},~~~
\widetilde{E}_n(\beta)=\beta^{1/5\,}\widehat{E}_n(\al(\beta))-\frac{2}{27\beta},
~~~\al(\beta)=\frac{1}{3\beta^{4/5}}\phantom{xxx}
\label{BARR_new}\ee
are easily obtained by composing the following analytic translation  and the dilation \cite{Combes} 
\be x\ra x+i/{3\sqrt{\beta}}\,,\qquad\quad x\ra \beta^{-1/10}x\label{TransDil}\ee

Later on we will prove that the eigenvalues $ \widehat{E}_n(\al)$ are real analytic for
$\al\in\mathbb{R}$ in a neighborhood of the origin. Through (\ref{RHT_new}) and 
(\ref{BARR_new}) the levels $E_n(\hbar)$ are analytic continuations of the perturbative 
levels $\widetilde{E}_n(\beta)$ and extensible as many-valued functions to the sector 
of the $\hbar$ complex plane
\be \mathbb{C}^0=\{\hb\in \mathbb{C}; \,\hb\neq 0, \,|\arg(\hb)|<\pi/4\},\label{SIH}\ee
We can now formulate in very simple terms the crossing selection rule. The two positive levels 
$E_{m^\pm}(\hb)$,
${m^\pm}={2n+(1/2)\pm (1/2)}$,  defined for sufficiently high $\hbar$ undergo a 
crossing at a value $\hb>\hbar_n$ and become the two complex conjugate levels $E_n^\pm(\hb)$ 
defined in
(\ref{isospettrobeta}) for $\hb<\hbar_n$. The 
corresponding states $\psi^\pm_{n}(\hb)$ are $PT$-conjugated. We will refer to   $E^c_{n}>0$  as 
to the limit level at $\hb=\hb_n$ and to $\psi^c_{n}$ as to the corresponding $PT$-symmetric  
state.

This process is possible because of the instability of a single node of $\psi_{m^+}(\hb)$ and 
the instability of the $PT$-symmetry of both the states $\psi_{m^\pm}(\hb)$.
More explicitly, 
the crossing rule is given in terms of the analytic continuations. 
The two functions $\,E_{m^\pm}(\hb)\,$,   holomorphic for large $\,|\hb|\,$, 
are analytically continued along the positive semi-axis for decreasing $\hb$  by passing above 
the singularity at $\hb=\hbar_n$ and becoming 
respectively the two levels $E_n^\mp(\hb)$   
for small   $\hb>0$. 
Thus the Bender-Wu singularities are square root branch points.
As  $n\ra\infty$, we have the limits  $\hb=\hb_n\ra 0$  and $E^c_{n}\ra E^c\geq0$, where $E^c$ is 
an instability point of the set of the Stokes lines supposed unique.

We give a brief summary of the content of the following sections where all the statements will be 
rigorously proved. 
In Sec. 2 we study the levels and the states for small parameter $\hb$. In particular, we
show  the stability of the nodes  in one of the half planes $\pm\Re z>0$.  In Sec. 3 we deal with 
the behavior of the levels and of the nodes for large values of $\hb$.
We prove a confinement of the nodes and the positivity of the spectrum, 
and the possible existence of only one  
imaginary node, which is the one becoming unstable at $\hb_n$. We also prove the selection rules of 
the crossings.
In Sec. 4 we show the semiclassical nature of the problem,  we give the 
quantization rules and we prove the local boundedness of the levels. We then determine the 
crossing  rules and we consider the Riemann surfaces of the levels in a neighborhood
of the real axis of $\hb$. In Sec. 5 we give the conclusions and suggest possible extensions to 
complex values of the parameter $\hb$.

\section{Confinement of the escape line  at $\boldsymbol{E\!=\!E_n^\pm(0)}$ and the zeros of 
$\boldsymbol{\psi^\pm_n(z,\hb)}$  
for small ${\boldsymbol{\hb}}\,$}

Let us consider the Stokes complex in our case. For any energy $E\in\mathbb{C}$, a  Stokes 
line is 
defined by a starting point, one of the turning points called  
$I_\pm$ and $I_0$.  At a point $z$ the Stokes vector  $ dz(z)$ satisfies the condition
\be -p_0^2(E,z)\,dz^2(z)>0,\,\quad\,p_0^2(E,z)= V(z)-E\label{STOKES1}\ee 
 where the choice of the sign is consistently defined in all the complex plane. In such a 
 way we give an orientation to the Stokes lines.
We are interested  in two
particular Stokes lines: the internal one (hereafter called the\emph{ oscillatory range} $\rho(E)$) 
and the 
exceptional one (called the \emph{escape line} $\eta(E)$)\cite{GI}. Their union 
$\rho(E)\bigcup\eta(E)$ 
is the \emph{union of the classical trajectories} $\tau(E)$. For later use it
is also convenient to introduce the following notations:
\be \mathbb{C}^\pm=\{z\,,\,\Re(z)\gtrless 0\}\,\qquad\quad
 \mathbb{C}_\pm=\{z\,,\,\Im(z)\gtrless 0\}\,
\label{Cpm}\ee 
 
With the definitions (\ref{Hpm}) of 
$E_0$ and $c$ we prove 
\medskip

\begin{lemma} 
	\textit{
		The escape lines $\eta(E)$ at the levels $E=\mp E_0$,  are in $\mathbb{C}^\pm$ 
		respectively, and 
		the $\rho(\mp E_0)$are the stationary points $x_\mp\in\mathbb{C}^\mp$. }  
	\label{lemma11}
\end{lemma} 

\textbf{Proof.} We fix $E=-E_0$. The case $E=E_0$ is completely analogous.
The turning point  $I_0=2/\sqrt{3}\in\mathbb{C}^+$ is the
 starting point  of the oriented exceptional Stokes line.
 The two turning points $I_\pm$ coincide and we 
have $\rho (E)=I_+=I_-$. Using the variable $w=y-ix$, 
the condition (\ref{STOKES1}) for the Stokes field becomes
\be -p_0^2(E,w)\,dw^2=(w^3+w+E)\,dw^2>0\label{STOKESC}\ee
For $w=-iI_0+\delta,$ at the first order in $\delta$ we have
\,$p_0^2(-iI_0+\delta)\,\delta^2\sim\, 3\delta^3<0\,
.\,$
Hence $\delta^3<0$ and $\arg\delta=\pm \pi/3.$ In the $z$ plane, $z=I_0+i\delta$ with $\arg 
(i\delta)=(\pi/2)\pm\pi/3$.
By the choice   $\arg (i\delta)=(\pi/2)+\pi/3=5\pi/6\,$, we
obtain the oriented exceptional Stokes line $\eta (-E_0)$.
We know that in the $z$ plane this line is asymptotic to the imaginary axis  at $+i\infty$.
For large $y>0$ the behavior of the action integral is
\be S(w)=\int_{-iI_0}^w \sqrt{-p^2_0(E,w)}\,\,dw(w)~\sim~ S_a(w)=(2/5)w^{5/2}+w^{1/2}+E w^{-1/2} 
~~\label{asiS}\ee 
Thus, if $w=w(y)=y-ix(y)\,$ and $\,x(y)\ra 0\,$ as $\,y\ra\infty\,,$ we have
\be S_a(w(y))=(2/5)\,y^{5/2}+y^{1/2} -iy^{3/2}\bigl((y^2+1/2)\,x(y)-\Im E\bigr)\,\,\ee 
and $\Im S_a(w(y))=0$ implies
\be x(y)\sim c/(y^2+1/2)\,.\label{ASST}\ee
Therefore the escape line $\eta(-E_0)$  stays  in $\mathbb{C}^+$. On the imaginary axis,  the 
vectors 
$\pm dw(y)$ are determined by the condition
\be p_0^2(E,y)\,dw^2(y)<0,\,\qquad\,p_0^2(y)=-(y^3+y)+E.\label{STOKES}\ee 
We consider the regular field of velocities on the imaginary axis letting  $dy>0$.
We make explicit the two conditions (\ref{STOKES}):
\begin{eqnarray}
&{}&\Re(p_0^2(E,y)\,dw^2(y))=(y^3+y)\,(dx^2-dy^2)+2\,c\,dx\,dy<0{\vrule height1.0em width0em 
depth1.0em}\cr
&{}&\Im(p_0^2(E,y)\,dw^2(y))=(y^3+y)\,2\,dy\,dx-c\,(dx^2-dy^2)=0.
\nonumber
\end{eqnarray}
Substituting the equality into the inequality we find 
$$(2/c)\,\bigl((y^3+y)^2+c^2\bigr)\,dx\,dy<0, $$
As $dy>0$ we get $dx<0$. Moreover, as the field of vectors $dw(y)$ is regular on the 
imaginary axis, the oriented line $\eta(E_0)$ could exit but not come back to $\mathbb{C}^+$, 
so that it stays always in $\mathbb{C}^+.$
\hfill $\square$
\medskip

The following useful result is a  consequence of the Lemma \ref{lemma11} and the exact 
semiclassical theory (See  \cite{ GI}, Theorem 1).

\begin{cor}
	Consider  the zeros of $\psi_n^\pm(z,\hb)$  with energy $E=E^\pm_n(\hb)$ and
	small $\hb.$ In the limit $\hb\ra 0^+$, $E^\pm_n(\hb)\ra\pm E_0$,  all the $n$ nodes 
	go to $x_\pm\in\mathbb{C}^\pm$ and all the other zeros go to $\eta(\pm E_0)\in\mathbb{C}^\mp$.
	\hfill $\square$ 
	\label{lemma_new}
\end{cor} 

We now study the behavior of levels and states in the semiclassical limit.

\begin{lemma} For any $n\in \mathbb{N},$ there exists $\hb_n>0$ such that 	
	for $0<\hb<\hb_n$ there are complex conjugate levels $E_n^\pm(\hb)$ whose
	corresponding states $\psi_n^\pm(\hbar,x)$ are $PT$-conjugated
	\be\psi_n^-(\hbar)=PT\psi_n^+(\hbar).\,\,\label{PTSS}\ee 
	Both the corresponding entire functions $\psi^\pm(z)$  have $n$ nodes respectively tending to 
	the points  $x_\pm\in\mathbb{C}^{\pm}$ as
		$\hb\ra 0^+$. Their  kernels in $\mathbb{C}$ are $P_x$-conjugated, namely 
	$\ker\psi_n^-(z)=P_x\ker\psi_n^+(z)$ where $P_xf(x+iy)=f(-x+iy)\,.$
	\label{lemma5}
\end{lemma} 
\textbf{Proof.$\,$} The isospectrality of $H_\hbar^\pm$ and $\widehat{H}_{\beta^\pm(\hbar)}$ has 
already established in (\ref{Hpm})-(\ref{isospettrobeta}). Let us only add that for positive $\hbar$
the parameters $\beta^\pm(\hbar)$ are not in in the complex plane cut along the negative axis,
$\mathbb{C}_c=\{z\in\mathbb{C};\,\,z\neq 0,\,\,|\arg z|<\pi\}$. This means that just the 
results of  \cite{GM} are not sufficient, but we also need some of the results of \cite {C}.
In particular we use the fact that there exists a $b_n>0$ such that the perturbative level $ 
\widetilde{E}_n(\beta)$ admits analytic continuations in the open disks of radius $b_n$ with 
centers at $\exp(\pm i\pi)\,b_n$ respectively. The perturbation theory yields that  the 
semiclassical behavior of the levels is
\be  E_n^\pm(\hb)=\pm E_0  + \hb c^\pm (2n+1)+O(\hb^{2})\,.\label{0B}\ee
We prove that the corresponding states $\psi_n^\pm(\hb)$ are $PT-$conjugated 
for a suitable choice of the phase factors. Indeed $PT$ is a bounded involution.
Therefore, from the relation $H\psi^+=E^+\psi^+$ we get
$(PT\,H\,PT)\,(PT\psi^+)=H(PT\psi^+)=\bar E^+(PT\psi^+)=E^-(PT\psi^+),$
which implies  (\ref{PTSS}) since the spectrum is simple.
Moreover, the set of the zeros of $\psi^-$ is the reflexion with respect of the
imaginary axis of the set of zeros  of $\psi^+$, or $\ker(\psi^-)=P_x\ker(\psi^+).$ It is
relevant to notice that in the perturbation theory of $H_\beta$ (\ref{Hbeta}), all the nodes of
$\widetilde{\psi}_n(\beta^\pm(\hb))$ have a semiclassical limit in 
$\rho (2n+1)=[\,-\sqrt{2n+1},\sqrt{2n+1}\,],$ while the corresponding nodes of the semiclassical
functions $\psi_n^\pm(\hb)$ go to the stationary points $x_\pm$ respectively.\hfill$\square$
 \medskip

We next prove that the zeros $\psi_n^\pm(\hb)$ are stably confined in $\mathbb{C}^\pm$ 
respectively, so that such zeros coincide with the nodes tending respectively 
to $x_\pm$ as $\hb\ra 0$. This shows that no crossing exists between levels of the same set 
$\{E_n^-(\hb)\}$ or $\{E_n^+(\hb)\}$. Since  $E_n^\pm(\hb)$ are complex conjugated 
for small $\hbar$, the levels  $E_n^-(\hb)$ and  $E_n^+(\hb)$ will cross at $\hb_n\,,$ where they 
become real.  
 \medskip 

\begin{lemma} 
	Let $E=E^\pm_n(\hb)$  be the non-real levels at $0<\hb<\hbar_n$. The corresponding states  
	$\psi_n^\pm(z)$  are non vanishing on the imaginary axis.   \label{lemma6} 
\end{lemma} 

\textbf{Proof.} Le $E$ be one of the non-real levels and $\psi(z)$ the corresponding state.
Let $\phi(y)=\psi(iy)\,,~y\in\mathbb{R}\,,$ be the eigenfunction on the imaginary axis.
With $w=y-ix$, for a fixed $x$ we define the Hamiltonian 
\be H^r_\hb(x)=-\hb^2({d^2}/{dy^2})-w^3-w\label{TransH}\ee 
having a level $-E\,.$ The corresponding
translated state,  $\phi_x(y)=\phi(w)$, has the well known asymptotic behavior for large $y$ 
\cite{S}
\be\phi_x(y)= 
\frac{C\bigl(1+O(y^{-1/2})\bigr)}{\sqrt{p_0(E,w)}}\cos 
\bigl(\frac{S_a(w)}{\hb}+\theta\bigr),\,\quad 
\label{ABS}\ee 
where $C>0\,$, $\theta\in\mathbb{R}\,,\,$ $p_0(E,w)$ is defined in (\ref{STOKESC}) and $S_a(w)$ in 
(\ref{asiS}). We have
\be |\phi(y)|^2 = |\phi_0(y)|^2=O(|y|^{-3/2})\,\quad \mathrm{as}\quad  y\ra+\infty\,.
\label{ABphi}\ee

We now consider the Loeffel-Martin formula for $\phi(y)$, producing the same result of the law
of the imaginary part of the shape resonances:
\be \hb^2\Im (\bar\phi(y)\phi'(y))=-\Im E\int_y^\infty|\phi(s)|^2ds, \forall 
y\in\mathbb{R},\label{NOZEROIMAX}\ee
Due to the asymptotic behavior (\ref{ABphi}) the integral exists  and is finite. Therefore 
$\psi^\pm_n(z,\hb)$ is not vanishing for $z$ on the imaginary axis.\hfill$\square$

\begin{lemma} 
	Let $E=E^\pm_n(\hb)$ and $\psi_n^\pm(z)$ as above.  The large zeros $Z_j^\pm$ of 
	$\psi_n^\pm (z)$ are in the half planes $\mathbb{C}^\mp$ and their nodes  are stable in 
	$\mathbb{C}^\pm$
	respectively.
	\label{lemma5bis} 
\end{lemma}
\textbf{Proof.} For $y\ra\infty$, $x(y)\ra 0$,  we get from (\ref{ABS})  the asymptotic condition  
$$\Im \bigl({S_a(w(y))}+\hb\theta\bigr)\sim -\bigl(y^{3/2}+1/2\bigr)\,x(y)+\Im E+\hb y^{1/2}\, 
\Im\theta\, =0$$
If $\Im \theta=0,$ the large zero $Z_j=x(y)+iy$  has an asymptotic behavior with
\be x(y)\sim ({\Im E_n^\pm+\hb y^{1/2}\,\Im \theta})\,/\,({y^{2}+1/2}),\,\,\label{SI1}\ee
These asymptotic behaviors are imposed by  the continuity of the zeros and their 
impossibility of crossing  the imaginary axis by Corollary \ref{lemma_new}. This proves the 
stability  of  the zeros (nodes) in $\mathbb{C}^\mp$ respectively.
At the limit of $\hb\ra \hb_n^-$ the energies $E_n^\pm(\hb)$ become positive and the large zeros 
$Z_j^{\pm}$ become  imaginary.
\hfill$\square$
\medskip

From \cite{ GI}, the Corollary \ref{lemma_new},the continuity of the nodes 
and Lemma \ref{lemma6} describing the barrier on the imaginary axis, we have

\begin{cor} 
Let $E=E^\pm_n(\hb)$ and $\psi_n^\pm(z)$ as above. For small $\hbar$  all the $n$ nodes are in a 
neighborhood $U_\pm\subset\mathbb{C}^\pm$ of $x_\pm$.\hfill$\square$
\end{cor}

We finally have an analyticity result for the levels $E_n^\pm(\hb)$.

\begin{prop}  	 
	The two functions $E_n^\pm(\hb)$ are analytic for $0<\hb<\hb_n.$
	The two levels $E_n^\pm(\hb)$ and the two states 
	$\psi_n^\pm(\hb)$ coincide at the crossing limit $\hb\ra\hb_n^-\,.$
	The limit level $E_{n}^c$ is positive. The limit state  $\psi_{n}^c(z)$  is $P_xT$-symmetric 
	and has $2n$ non-imaginary 
	zeros. The large zeros are imaginary.
	\label{lemma7}
\end{prop} 
\textbf{Proof.}  
According to Lemma \ref{lemma5} and Lemma \ref{lemma5bis}, for $\hb<\hb_n$ the $n$ nodes of the 
two states  $\psi_n^\pm(\hb)$,  are the only zeros   in $\mathbb{C}^\pm$, respectively. 
Since the states $\psi_n^\pm(\hb)$ are the only ones having $n$ zeros in 
$\mathbb{C}^\pm$ 
the  function $E_n^\pm(\hb)$ are analytic. From the relation (\ref{PTSS}) for $\hb<\hbar_n$ and 
the limit
$\psi_n^\pm(\hb)\ra\psi_{n}^c$ as $\hb\ra \hbar_n^-$ we get $\psi_{n}^c=PT\psi_{n}^c.$ 
As the 
states $\psi_n^\pm(\hb)$ have only $n$ zeros in $\mathbb{C}^\pm$ and 
at the limit $\hb\ra \hbar_n^-$ these zeros cannot diverge or become 
imaginary,  the limits of the $2n$ non-imaginary  zeros of both the state $\psi^\pm(\hb)$ are all 
the non-imaginary zeros 
of the limit state $\psi_{n}^c.$ \hfill $\square$

\bigskip 

\section{Analysis of levels and nodes for large $\hb$}

We have already stated in the Introduction that level crossing comes from looking at the behavior 
of the levels for small and large $\hb$. We have also described in (\ref{KHH})-(\ref{TransDil}) the
appropriate scaling for dealing with large values of $\hb$ or $\beta$, corresponding to  small 
values of $\al$. 

We are now going to prove  a confinement of the 
zeros of $\widehat{\psi}_n(\al)$, for small $\al$, in two regions. We define nodes the zeros 
confined in one 
of these regions by identifying them  with the nodes of the states 
$\widetilde\psi_n(\beta)$ for large $\beta$. We find it useful to introduce the two disjoint sets
\be
&{}&\mathbb{C}_\rho=\{z=x+iy,\,\,y<0,\,|x|<-\sqrt{3}\,y\}\,\,
\subset\,\,\mathbb{C}_-\phantom{xxxxx} \spazio{0.8}\cr
&{}&\mathbb{C}_\eta=\{z=x+iy,\,\,y>0,\, |x|<\phantom{-}\sqrt{3}\,y\}
\,\,\subset\,\,\mathbb{C}_+\label{CRE}\ee 

\begin{lemma} 
	{The $m$ nodes of  
	$\widehat\psi_m(\al)$ for small $|\al|$ are confined in  $\mathbb{C}_\rho$ and correspond to 
	the 	nodes  of $\widetilde \psi_m(\beta)$ in $\mathbb{C}_-$.
		The other zeros of $\widehat\psi_m(\al)$ are in $\mathbb{C}_\eta$. The function $\widehat 
		E_m(\al(\beta))$ 	is real analytic and coincides with
		$\widetilde E_m(\beta)$ by  $(\ref{BARR_new})\,.$
	}\label{lemma_ini2}
\end{lemma} 
\textbf{Proof.}
Define the translated operator $\widehat{H}_{\al=0}$ by $x\ra x+iy\,,$ 
\be \widehat{H}_{\al=0,\,y}=-\,(d^2\!/dx^2)+V_y(x)\,,\qquad V_y(x)=y\,(y^2-3x^2)+ix\,(x^2-3y^2)
\label{(TransAlpha)}\ee
Apply the Loeffel-Martin method \cite{LM} to the state $\psi=\widehat\psi_m(\al=0)$ with energy 
$E=\widehat{E}_m(\al=0)>0$, for  $\,\pm x\geq \sqrt{3}\,|y|\,:$
\begin{eqnarray}
&{}& -\Im \,[\overline{\psi}(x+iy)\pa_x\psi(x+iy)]=\int_{x}^\infty\Im 
V_y(s)\,|\psi(s+iy)|^2ds=\cr
&{}&\phantom{xxx} \int_{x}^\infty(s^2-3y^2 )
s|\psi(s+iy)|^2ds=-\int^{x}_{-\infty}(s^2-3y^2 ) s|\psi(s+iy)|^2ds\neq 0
\nonumber
\end{eqnarray}
For $\al=0$ the nodes are thus rigorously confined in $\mathbb{C}_\rho.$ 
The confinement extends to $\al>0$.  Since the $m$ zeros of $\widehat{\psi}_m(\al)$ on 
$\mathbb{C}_-$ are stable for  $\al\ra+\infty\,,$ 
they are nodes by definition. 
By (\ref{TransDil}) the set
$\mathbb{C}_-$ is invariant for positive dilations  in the limit of 
infinite $\beta$ and for $\beta>0$ the $m$ nodes of $\widetilde\psi_m(\beta)$  are in 
$\mathbb{C}_-$  \cite{GM}. 
By (\ref{BARR_new}) the level $\widehat E_m(\al)$ for small $|\al|$ connects 
the perturbative level 
$\widetilde 
E_m(\beta)$ with the level $E_m(\hb)$ for large  positive $\beta$ and $\hb$ respectively.
The state $\widehat{\psi}_m(\al)$ is defined by the number $m$ of its zeros in 
$\mathbb{C}_-$ which,  by Lemma \ref{lemma_ini2}, are identified as nodes.
Since by continuity the number of nodes  of $\widehat{\psi}_m(\al)$ is stable for $0<|\al|<\es\,$
while, although unbounded, the scaling (\ref{dilalambda}) is regular and phase preserving, 
the function $\widehat{E}_m(\al)$ is real analytic for all $\al>0$ due to
the results of \cite{GM} and to (\ref{RHT_new}). This property is stable for small $-\al> 0\,$,  
up to the first crossing and the function $E_m(\hb)$  is real  analytic by (\ref{RHT_new}) and 
(\ref{BARR_new}). Finally the $m$ nodes of $\psi_m(\hb)$ for large positive $\hb$ are 
all its zeros in $\mathbb{C}_\rho$ as well as in  $\mathbb{C}_-$.\hfill $\square$

\begin{cor}
	For large $\hb$ the level $E_m(\hb)$ exists and  is  positive.
	\label{lemma3.3}
\end{cor} 

\textbf{Proof.} The analytic level  $\widehat{E}_m(\al)$, with corresponding normalized state  
$\widehat{\psi}_m(\al)$, has positive real part due to the positivity of $\,-\,(d^2\!/dx^2)\,$:
$$\Re \widehat{E}_m(\al)=\Re 
\langle\widehat{\psi}_m(\al),\widehat H_\al\,\widehat{\psi}_m(\al)\rangle
=\langle\widehat{\psi}_m(\al),-\,(d^2\!/dx^2)\,\widehat{\psi}_m(\al)\rangle\,>\,0,$$
The result follows from (\ref{RHT_new}).\hfill $\square$ 
\medskip

We now establish some properties of the $PT$-symmetric states which will be useful later on. 
In analogy to (\ref{(TransAlpha)}) we consider the translated operator
\be H_{\hb,y}=-\hb^2\,(d^2\!/dx^2)+V(x+iy)\label{TransHbar}\ee
with $V$ as in (\ref{RKHKC1}).

\begin{lemma} 
	A  $PT$-symmetric state  $\psi_y (x)=\psi(x+iy)$ of the translated operator $H_{\hb,y}$ has 
	even 
	real part and odd imaginary part. In particular it satisfies  at the origin the conditions  
	\be \Im\psi_y(0)=\Im\psi(iy)=0,\,\quad\, \Re\psi_y'(0)=\Re\psi'(iy)=0.\label{ICATO}\ee 
	\label{lemmaextra}
\end{lemma} 
\textbf{Proof.} Indeed if $\psi_y(x)=R(x)+iI(x)$ we have
$$\phantom{XXXXXXX}PT(R(x)+iI(x))=R(-x)-iI(-x)=R(x)+iI(x)\phantom{XXXXXX}\square$$  
\medskip

Let us recall the criterion for the nodes in the case of a positive level $E_m(\hb)$. 
A zero $Z_j(\hb)$ of $\psi_m(\hb,z)$ is a node  if,   continued to a  parameter $\hb'>\hb$ large 
enough, $Z_j(\hb')$ belongs to $\mathbb{C}_-$, \emph{i.e.}  $\Im Z_j(\hb')<0$. Let also recall that 
an imaginary  zero $Z_j(\hb)$ of the state $\psi_m(\hb,z)$ stays imaginary for any $\hb'>\hb$, 
because of the $P_x$-symmetry of its kernel and the simplicity of the spectrum.

\begin{lemma} The level $E_m=E_m(\hb)$ exists positive for $\hb$ large enough, 
	with the 	corresponding \textit{PT}-symmetric state 
	$\psi_m(x)=\psi_m(x,\hb)$.
	There is an alternative:
	\smallskip 
	 
	$(a)$ the absence of imaginary nodes of the function $\psi_m(z)$,
	\\$(b)$ 
	the existence of only one imaginary node of the function $\psi_m(z)$.
	\smallskip
	 
	The second case is possible if and only if $m$ is odd.
	\label{lemma4}
\end{lemma} 

\textbf{Proof.} For $\hb$ large enough the Hamiltonian (\ref{RKHKC1}) admits a positive level  
$E_m=E_m(\hb)$  with eigenfunction $\psi_n(z)=\psi_n(z,\hb)$. We 
consider  the  Hamiltonian (\ref{TransH}) on the imaginary axis, $x=0$ or $w=y$, and we observe that
$H^r_\hb=H^r_\hb(x=0)=-H_\hb$ is real. The  eigenfunction  $\phi_m(y)=\psi_m(iy)$ of $H^r_\hb$ 
is also real by the conditions (\ref{ICATO}) at the origin. $H^r_\hb$  and $-H_\hb$ have the same  
spectrum and  $-E_n=-E_n(\hb)<0$ is one of its eigenvalues. Therefore for large positive $y$
the solution $\phi_m(y)$ is the function $\phi_{x=0}(y)$ given in (\ref{ABS}).
For large $\,-y>0\,$ the solution $\phi_m(y)$  is a real  combinations  of the two fundamental 
solutions \cite{S} and reads
\be\phi_m(y)= \frac{C'\bigl(1+O((-y)^{-1/2})\bigr)}{\sqrt{p_0(E,y)}}
\,\Bigl(\exp \bigl(\frac{S_a(-y)}{\hb}\bigr)
+a\exp \bigl(-\frac{S_a(-y)}{\hb}\bigr)\Bigr),\,\,\label{AAPI1}\ee 
with $C'>0\,$,  $a=a_m(\hb)\in\mathbb{R}\,$ and $p_0(E,y)=\sqrt{-y^3-y+E}\,.$

For  $[m/2]=n\in\mathbb{N}$ and $\hb\geq \hbar_n$ both functions $\psi_m(z)$ have  $n$ 
nodes on both half-planes $\mathbb{C}^\pm$. They are distinguished by the number
of imaginary nodes. If we define the the complement of the escape line in the imaginary axis,
\be \eta^c(E)=\{z=iy,\, -\infty<y<y_0\},\label{CBOIA}\ee
where $y_0=-i\,I_0\,$ and $\,I_0\,$ is the imaginary turning point, then 
the crossing process for $\hb\geq \hbar_n$  can be studied by looking at the behaviors
$\psi_m(z)$ with energy $E_m$ on the semi-axis $ \eta^c(E_m)\,$. We therefore consider the 
behavior of the two states $\phi_m(y)$ in an open interval $A\subseteq \eta^c(E_m)$ for large $\hb$
and we see that a state is concave when it is positive and convex when negative.
Since we can consider $\phi_m(y)$ positive decreasing for $y \ll y_0\,,$
only two possibilities are admitted:
\smallskip

$(a)$ there exists a single zero on $\eta^c(E_m)\,;$\\
$(b)$ no zero exists on $\eta^c(E_m)\,.$
\smallskip

According to {\rm{Lemma \ref{lemma_ini2}}}, for large positive $\hb$ an imaginary node of a state 
$\psi_n(z,\hb)$ is in $\mathbb{C}_-\,$. As $y_0>0$ when $E>0$ an imaginary node
should lie in the intersection of  $\mathbb{C}_-$ with the imaginary axis, contained in 
$\eta^c(E_m)\,.$\hfill$\square$

\begin{cor}  	
	Assume   $\hb>\hbar_n$ and let $\psi_m(\hb)$  be a generic state tending to 
	$\psi_{n}^c$ for $\hb\ra \hbar_n^+.\,$ Then  the non-imaginary zeros of $\psi_m(\hb)$  are 
	exactly  $2n$  and they are  stable at the limit $\hb^+_n$.  Since at most there exists  
	one imaginary node, the number $m$ of the  nodes of  $\psi_m(\hb)$ is not greater than  
	$m^+=2n+1$.
	\label{lemma8}
\end{cor} 
\textbf{Proof\phantom{..}}  By {\rm{Lemma \ref{lemma4}}} when $\hb>\hb_n$ the levels $E_m(\hb)$ 
are positive and the states $\psi_m(\hb)$ are \textit{PT}-symmetric. Since the spectrum is simple 
and the nodes are symmetric, a zero on the imaginary axis cannot leave it and a non imaginary zero 
cannot become purely imaginary. A non imaginary zero of the state  $\psi_m(\hb)$, however,  can
go to infinity moving along a path having the imaginary axis as asymptote at infinity \cite{S, GM}.
Moreover  at a fixed $\hb>\hb_n$ the state  $\psi_m(\hb)$ has the behavior  (\ref{ABS})  so that it 
is non vanishing for large $y$ and small $|x|\neq 0$. We can therefore conclude that the large 
zeros  are imaginary and the non imaginary nodes are stable. 
The non imaginary zeros of the two states $\psi_m(\hb)$ as well as of the 
limiting state $\psi_n^c$ are $2n$ and by Lemma \ref{lemma4} there exists 
at most one zero on the imaginary axis. Thus the number of nodes  is $\,0\leq m\leq 2n+1$.
Due to the independence of the two states $\psi_m(\hb)$ we actually have two different numbers $m$,
one not greater than $2n+1$ and the other not greater than $2n$.\hfill$\square$ 
\medskip

Collecting all the previous results we can finally prove the
\begin{thm} 	
	For each $n\in\mathbb{N}$, there exists a crossing parameter $\hbar_n$. Two levels 
	$E_{m^\pm}(\hb)$,  $m^\pm=2n+(1\pm 1)/2$,  are defined for $\hb>\hbar_n$ and 
	two levels $ E_n^\pm(\hbar_n)$ for $\hb<\hbar_n$.  The two pairs cross at $\hbar_n.$
	Both the states $\psi_{m^\pm}(z,\hb)$ have a $P_x-$symmetric set of $\,2n$ 
	non-imaginary nodes. Only $\psi_{m^+}(z,\hb)$ has an imaginary node.
	\label{thm1} 
\end{thm} 

\textbf{Proof\phantom{..}} Before giving the proof an observation is in order. Actually
we do not prove the uniqueness of this crossing and the possible existence of a 
next pair of anti-crossing and crossing is left open. It could happen that the two levels 
$E_{m^\pm}(\hb)$, for $\hb'_n>\hb>\hbar_n$, and the two levels $ E_n^\pm(\hbar)$ for 
$\hb>\hbar'_n$, cross at $\hbar'_n\,.$  Moreover  $E_{m^\pm}(\hb)$, for 
$\hb>\hbar''_n$, and 
$ E_n^\pm(\hbar_n)$, for $\hb<\hbar''_n$, cross at $\hbar''_n\,.$  For simplicity, 
we disregard this possibility.	\\
Let us now proceed with the proof of the theorem.
The non reality of $E_n^\pm(\hb)$ for small $\hb$ and positivity of $E_m(\hb)$ for large $\hb$
necessarily yield the existence of crossings. Seen from $\hb\leq \hbar_n$ the crossing
occurs when the two levels $E_n^\pm(\hb)$ become real. Considering the pairs of positive levels, of 
the kind called $E_m(\hb)$, obtained by the crossing 
at $\hb_n$, only  the pairs of numbers $m$ equal to $m^\pm$ are compatible 
with the uniqueness of such levels for large $\hb$, namely only the sequence of pairs, 
$\{(E_{2n}(\hb),E_{2n+1}(\hb))\}_{n=0,1,...}$
corresponds  exactly to the  
sequence  
$\{E_m(\hb)\}_{m=0,1,...}$
we have for large $\hb$.
As a consequence we also have that  the non-imaginary zeros of the states $\psi_m(\hb)$ are the 
non-imaginary nodes.\\
At the crossing, the   imaginary  node of the state 
$\psi_{m^+}(\hb_n)$ 
coincides with the lowest imaginary zero of $\psi_{m^-}(\hb_n)$. 
The crossing between the levels $E_{m^\pm}(\hb)$ is possible because of the 
stability of the $2n$ non-imaginary nodes of both the entire functions $\psi_{m^\pm}(z,\hb)$ and 
the instability of 
the imaginary node of the functions $\psi_{m^+}(z,\hb)$. Because of the $P_xT$-symmetry of both the 
states $\psi_{m^\pm}(z,\hb)$, they have $n$ nodes in both the half-planes. If we continue the state 
$\psi_{m^+}(z,\hb)$ along a path in the $\hb$ complex plane, coming and returning to a $\hb>0$, 
large enough, turning around $\hb_n$,  at the end we get the state with $m^-$ zeros in the 
half-plane $\mathbb{C}_-$ of $z$, without the imaginary one. 
Finally, by continuing  to $\hb<\hb_n$ the $n$ nodes of both the states 
$\psi_{m^\pm}(z,\hb)$
in $\mathbb{C}^+$ ($\mathbb{C}^-$) we obtain the nodes of  $\psi_n^+(z,\hb)$ 
($\psi_n^-(z,\hb)$).\hfill $\square$ 

\begin{rems}
	{\rm{$(i)$	
	For large $\hb$ all the zeros in the upper half-plane  are imaginary.
	This statement strengthens  the confinement of the zeros of $\psi_{m}(z,\hb)$ for large $\hb$ 
	obtained above. 
	It ensues from the 
	result that all the non-imaginary zeros are nodes, and all the nodes are in the lower 
	half-pane for large $\hb\,.$
	\medskip
	
$(ii)$ We have seen that the states $\widetilde{\psi}_n=\widetilde{\psi}_n(0)$ of 
$\widetilde{H}_\beta$ at  fixed $\beta=0,$ 
have definite parity: $P\widetilde{\psi}_n=(-1)^n\widetilde{\psi}_n$. This means that 
$|\widetilde{\psi}_n|^2$ 
is $P$-symmetric, and the expectation value of the parity is 
$\,\langle\widetilde{\psi}_n,P\widetilde{\psi}_n\rangle=(-1)^n$. We recall that the state 
at the  crossing, $\psi_{n}^c=\psi_n^\pm(\hbar_n)$  has vanishing average value of
the parity,
$\langle\psi_{n}^c,P\psi_{n}^c\rangle=0,\,$ so that it is totally $P$-asymmetric in the sense 
that $\psi_{n}^c$  is orthogonal to $P\psi_{n}^c$ \cite{GG}.	
}}
	\label{rem1}
\end{rems}

\section{Boundedness of the levels, quantization and selection rules}

In this section we examine further properties of the levels in connection with the quantization and
the selection rules.\\
The levels are always semiclassical and are given by some semiclassical quantization rules 
excluding the divergence of the level. The semiclassical nature of the problem is made clear 
using the dilations. By a regular scaling $x\ra \lambda x,$ $\lambda =1/\sqrt \delta>1$ we get the 
equivalent operator
\be \check{H}_k(\delta)=-k^2\,(d^2\!/dx^2)+i(x^3-\delta x)\sim  \delta^{3/2} H_\hb\,,\qquad 
\,\,k=\hb\,\delta^{5/4}\label{KDH}\ee
where the new parameter $k$ vanishes, for any fixed $\hb$, as $\delta\ra 0\,.$
In this new representation the energy is $ \check{E}_m(k,\delta)=\delta^{3/2}E_m(\hb)\,.$ 
Fixing $\delta= 0$ and setting $k=\hb$ we reproduce the well known semiclassical operator \cite{BB}
\be\check H_\hb(0)=-\hb^2\,(d^2\!/dx^2)+ix^3\,
\label{CHECH}
\ee
 We recall that $\check H_\hb(0)$ is related by a 
 scaling to the operator  $\widehat{H}_{\alpha=0}$ studied above, its spectrum is positive and the 
 Stokes
 lines have the trivial dependence  
 $\tau(E)=E^{1/3}\tau(1)$ upon $E>0$, so that there are no critical energies. Thus,  
 we can consider a large enough scaling factor $\lambda$ in order to get the  parameter $k$,
 replacing $\hb$, as small as we want.

Bounded levels $E_n^\pm(\hb)$, $E_m(\hb)$ are obtained by two different quantizations. 
We recall that the nodes of a state $\psi$ are confined in  $\mathbb{C}^\pm$ according to whether 
its energy satisfies the condition $E\in\mathbb{C}^\mp$. Therefore for $\hb<\hb_n$ both levels
$E_n^\pm(\hb)$ satisfy the  unique conditions on the imaginary part and on the nodes of the states,
have no crossing and are analytic. At $\hb=\hbar_n$ the levels cross and become positive.
We have seen that there are two continuations of $E_n^\pm(\hb)$ from $\hb<\hb_n$ to $\hb>\hb_n$ 
and that the continuations of the 
corresponding states have $n$ nodes each one in the half planes $\mathbb{C}^\pm\,.$ There exist
two regular regions $\Omega^\pm\subset\mathbb{C}^\pm$ large enough whose boundaries  
$\,\gamma^\pm=\partial \Omega^\pm\,$ satisfy $\, P_x\gamma^+=\gamma^-\,$ such that the 
exact quantization conditions are
\be J^\pm(E,\hb)=
\frac{\hb}{2i\pi}\oint_{\gamma^\pm}\frac{\psi'(z)}{\psi(z)}dz+\frac{\hb}{2}=\hb \Bigl( 
n+\frac{1}{2}\Bigr), 
\label{CC}\ee
with $E=E_n^\pm(\hb)\in\mathbb{C}^\mp\,$ and  $\psi(z)=\psi_n^\pm(z,\hb)$. For
small $\hb\,$ and  bounded  $n\hb\,$, from (\ref{CC}) we get the semiclassical quantization
\be J^\pm(E,\hb)=
\frac{1}{2i\pi}\oint_{\gamma^\pm}\sqrt{V(z)-E}\,dz+O(\hb^2)=\hb \Bigl( n+\frac{1}{2}\Bigr), 
\label{CC1}\ee
where  $\gamma^\pm$ squeeze along both the edges of $\rho(E)$. At the critical value $\hb=\hbar_n$
the two quantizations (\ref{CC}) yield equal solutions $E_{n}^c$, $\psi_{n}^c\,.$ When 
$\hb>\hb_n\,,$
both (\ref{CC}) admit the two  solutions $E_m(\hb)$, $\psi_m(\hb)$, $[m/2]=n\,$ distinguished 
by the selection condition  $E_{2n+1}(\hb)>E_{2n}(\hb)\,$ 
compatible with the order of the levels $\widehat E_{2n+1}(0)> \widehat E_{2n}(0)$.
Thus, we have the boundedness and continuity of the functions $E_n^\pm(\hb)$, 
$\hb\leq\hb_n\,,$ becoming $E_{m^\pm}(\hb)$ for $\hb\geq\hb_n\,$ and both the
functions $E_{m^\pm}(\hb)$ are analytic in the hypothesis of maximal 
analyticity.

We now consider the semiclassical regularity of the levels $E_m(\hb)$ for  large $1/\hb$
and  $m=2n$ or  $m=2n+1\,.$
In particular, we expect to find positive semiclassical levels with $E>E^c=0,352268..$ 
\cite{DT} by a semiclassical quantization we consider here.
If  $\Omega_m\subset \mathbb{C}_-$ is large enough in order to contain all the $m$ 
nodes and $\Gamma_m=\partial \Omega_m\,,$ for a fixed $\hb,$ we have the exact  quantization 
rules
\be J_2(E,\hb) =
\frac{\hb}{2i\pi}\oint_{\Gamma_m}\frac{\psi'(z)}{\psi(z)}dz+\frac{\hb}{2}=\hb \Bigl( 
m+\frac{1}{2}\Bigr) 
\label{CC2}\ee
where $ E=E_m(\hb)\,,$ $\psi(z) =\psi_m(\hb,z)\,.$ For large $m$, small $\hb$, $m\hb$ bounded, we 
have
\be J_2(E,\hb)=
\frac{1}{2i\pi}\oint_{\Gamma_m}\sqrt{V(z)-E}\,dz+O(\hb^2)=\hb \Bigl( m+\frac{1}{2}\Bigr)\,.
\label{CC3}\ee
We expect a bounded limit of both $2n\hb_n\ra J_2^c$ and $E_n^c\ra E^c$ as $n\ra\infty.$ 
The peculiarity of $E^c$ is the instability of $\tau(E)$ and the connection of $\rho(E)$ 
and $\eta(E)$ at this point.
The quantization conditions (\ref{CC1}) and (\ref{CC3}) are compatible with the existence of 
$\lim_n E_{n}^c=E^c>0$ for $n\ra\infty$, $J_2^c=J_2(E^c,0)=2J(E^c,0)$ with  
$\Omega=\Omega^+\bigcup\Omega^-$. The localization of the nodes near  $\rho(E)$ for small 
$\hb$ and the localization of the other zeros near $\eta(E)$ implies this property of 
$\tau(E)$ at $E^c$.\\
We can now establish the local boundedness  of the levels in  the real axis.

\begin{lemma} 	 Each of the four  continuous functions 
	$E_n^\pm(\hb)$ for $\hb<\hbar_n$ and $E_{m^\pm}(\hb)$ for $\hb>\hbar_n$ is locally 
	bounded.
	\label{lemma10}
\end{lemma} 
\textbf{Proof\phantom{..}} Let $E(\hb)$ be one of the levels $E_n^\pm(\hb)$ for $\hb<\hbar_n$ 
with one of its continuations $E_{m^\pm}(\hb)$ for $\hb>\hbar_n\,$  and let $\psi(z)$ be the
corresponding state. Assume that the lemma is 
false and there is a divergence of $E(\hb)$ at $\hb^c\gg\hb_n\,.$ We rescale the Hamiltonian
$H_\hb$ by $x\ra |E(\hb)|^{1/3}\,x\,.$ Upon dividing by $|E(\hb)|$ we get the  
operator
\be -k^2\,(d^2\!/dz^2)+i\,z^3-i\,|E|^{-2/3}\,z -E/|E|\,,\qquad k=|E|^{-5/6}\,\hb
\nonumber
\ee
As $\hb\ra\hb^c$, so that $k\ra 0$, and neglecting the linear term in $z$, the semiclassical 
quantization reads
\be
\frac{1}{2\pi i}\oint_{\Gamma_m}\sqrt{iz^3-{E}/{|E|}}\,\,dz=k\,\Bigl(m+\frac{1}{2}\Bigr)
+O({k}^2),
\label{CC33}\ee 
where $\Gamma_m$ is the boundary of a region $\Omega_m\subset\mathbb{C}_-$ containing the
$m$ nodes of $\psi(\hb^c)\,.$ For $k\ra 0\,,$ (\ref{CC33}) could be satisfied only if
$E/|E|\ra 0$, obviously absurd.\hfill$\square$
\medskip

On the complex plane of the parameter $\hb$ consider now the sector $\mathbb{C}^0$ (\ref{SIH})
where the functions $E_{m^\pm}(\hb),$ are analytic with Riemann sheets $\mathbb{C}^0_{m^\pm}$ 
having a square-root-type singularity and a cut $\gamma_{n}=(0,\hbar_n\,]\,$ on the real axis.
We assume the inequality  
\be  E_{m^+}(\hb)>E_{m^-}(\hb),\qquad\,\,\hb>\hb_n\,,\label{C1}\ee 
the only one compatible with the order of the levels $\widehat E_{2n+1}(0)> \widehat E_{2n}(0)$.
We prove the following: 

\begin{thm} 
	The
	positive analytic functions $E_{m^\pm}(\hb)$  have the following behaviors at the edges of  
	$\,\gamma_{n}$
	\be E_{m^-}(\hb\pm i0^+)={E}^\pm_{n}(\hb)\,,\qquad\,E_{m^+}(\hb\pm 
	i0^+)={E}^\mp_{n}(\hb)\label{SON}\ee
	where ${E}^\pm_{n}(\hb)\ra\pm E_0$ as $\hb\ra 0.$ 
	\label{thm2}
\end{thm} 
	
	\textbf{Proof\phantom{..}} 
	In the hypothesis of unicity of the crossing (see the proof of Theorem \ref{thm1})
	for $\hb>\hb_n$ we admit the inequality   $E_{2n+1}(\hb)>E_{2n}(\hb)\,,$
	the only one compatible with the order of the levels $\widehat E_{2n+1}(0)> \widehat 
	E_{2n}(0)$. Since both the functions $E_{m^\pm}(\hb),$ have a square root 
	singularity at $\hb_n$ and $E_{m^+}(\hb_n+\es)-E_{m^-}(\hb_n+\es)=O(\sqrt{\es})>0$ for 
	small positive $\es\,,$ then
	$\,\pm\Im \,\bigl[\,E_{m^+}(\hb_n+\exp(\pm i\pi)\es)-E_{m^-}(\hb_n+\exp(\pm 
	i\pi)\es)\,\bigr]\,<\,0$ and $\mp\Im 
	E_n^\pm (h)>0\,.$  We necessarily  have 
	$$E_{m^+}(\hb_n+\exp\bigl(\pm 
	i\pi)\es\bigr)={E}^\mp_{n}(\hb_n-\es),\quad
	E_{m^-}\bigl(\hb_n+\exp(\pm i\pi)\es\bigr)={E}^\pm_{n}(\hb_n-\es)
	$$
	and the results extends to any $\es<\hb_n$. \hfill$\square$
	\medskip	
	
	\begin{rems} {\rm{	
		$(i)$ We can look at the crossing process following a path which starts from $\hb=0^+$,
		encircles the singularity $\hbar_n$ and comes back to  $\hb=0^+\,.$ At the beginning of
		the path the state $\psi_n^-(z,0^+)$ is mainly localized around $x_+$ and at the end
		turns into $\psi_n^-(z,0^-)$, mainly localized around $x_-\,.$ We can also look at a
		path beginning at a large $\hb$, going around $\hbar_n$ and returning to the initial 
		$\hb\,.$ If the initial state $\psi_{m^+}(\hb)$ is odd at the end it becomes the
		even state $\psi_{m^-}(\hb)$ and the imaginary node in the lower half plane is changed
		into the lowest zero on the positive imaginary axis.
		\medskip
			
		$(ii)$ It is possible that the Riemann sheet $\mathbb{C}^0_0$ of the fundamental level has 
		only 
		the square root branch point $\hb_0$ with the cut $\gamma_{0}=[0,\hb_0]$ on the real axis
		{\rm{\cite{DT}}}. From Theorem \ref{thm2} the discontinuity on the cut  $\gamma_{0}$ is
		$$ E_{0}(\hb+ i0^+)-E_{0}(\hb- i0^+)=2 \,i \,\Im 
		{E}^\pm_{0}(\hb)\,.$$
		The  function $E_0(\hb)$, analytic for large $|\hb|$, if continued to small $|\hb|$ while 
		keeping $\arg\hb=\pm\pi/4$ coincides by definition with  $E^\pm_0(\hb)$, respectively. 
		The 
		absence of complex singularities is compatible with  the identities on the edges of the 
		cut 
		$\gamma_{0}$
		 $$E_{0}(\hb\pm i0^+)=E^\pm_0(\hb\pm 
		i0^+).$$ 
	}}
		\label{rem3} 
	\end{rems}
We finally prove the theorem
\begin{thm} There exists  an  instability point, $E^c\geq 0\,$, of $\rho(E)\,.$ For 
	$n\ra\infty,$ we have 
	the limits $E_{n}^c\ra E^c\,,\,$  $2n\hb_n\ra J_2^c$  where $J_2^c=J_2(E^c,0)$ 
	as in $(\ref{CC3})$.
	\label{thm3}
\end{thm} 
\textbf{Proof\phantom{..}}  Since the eigenvalue problem of $H_\hb$ is semiclassical, the 
existence of the infinite crossings is possible if it exists  a critical point $E^c\geq 0$ of 
$\rho(E)$, which we assume to be unique. The existence of a critical point is due to the 
$P_x$-symmetry of 
$\rho\bigl(E_m(\hb)\bigr)$ for $\hb>0$ and $E_m(\hb)>0$ large, together with the symmetry 
breaking for small $\hb$. Actually at the limit $\hb\ra0^+$ we have at $E_n^\pm\ra\pm E_0$ 
and $\rho(\pm E_0)$ reduces to the points $x_\pm$, respectively. 
The breaking of $\rho(E)$ at $E^c\geq 0$ is possible only if $I_0(E^c)\in \rho(E^c)\,.$ 
In particular, we have  the $P_x$-symmetry breaking   at $E=E^c $ where $\rho(E)$
is a line touching  the turning points $I_\pm$ with $\Re I_\pm\neq 0$ and containing the 
point $I_0.$ 
Thus, the symmetry breaking of 
$\rho(E)$  implies its breaking at $E^c$ and its redefinition as one half of it containing 
only
a pair of turning points, 
$(I_0,I_+)$ or $(I_-,I_0)$.
Our  eigenvalue problem is always semiclassical and the change of 
the semiclassical regime is related to the instability  of the nodes used 
for the semiclassical quantization. 
Since it is possible to change representation by the scaling (\ref{KDH}), it is always possible 
to have a sequence of
parameters $k_n\ra 0$ together with
a sequence $\delta_n$. If a subsequence $\delta_{n(j)}$ vanishes as $j\ra\infty$, this is 
incompatible with the absence of crossings of the levels $\widehat E_m(\al)$ at $\al=0.$
Thus, we have the inequality of the sequence $\delta_n>\es>0$ for an $\es>0$ and $n>n_\es$. 
This means that also the original sequence vanishes, namely $\hb_n\ra 0$. Moreover, it is 
necessary that $E^c_n\ra E^c$ as $n\ra\infty,$ because in the semiclassical limit there is the 
instability of the nodes at $E=E^c$ only.
Thus, the sequence $E_n^c$ has the limit $E_n^c\ra E^c>0$ as $n\ra\infty$. But, due to
the semiclassical quantizations rules, we also have a bounded limit of  $2n\hb_n\ra 
J_2^c>0$  (\ref{CC3}), where $J_2^c=J_2(E^c,0)$ as $n\ra\infty.$ $\quad$\hfill$\square$

	\section{Conclusions}  We have proved the existence of a crossing for any pair of levels  
	$E_{m^\pm}(\hb)$, with  $m^\pm=2n+(1\pm1)/2)$,   giving the  pair of levels $E_n^\pm(\hb)$ for 
	smaller $\hb.$    In this \textit{PT}-symmetric model we see the competition of two 
	different effects: the conservation of the symmetry and the semiclassical localization of 
	the states. The semiclassical localization prevails for small $\hb$. This semiclassical 
	transition is impossible in families of selfadjoint Hamiltonians. Thus, these 
	\textit{PT}-symmetric models can be used to describe the appearance of the classical world 
	in  non isolated systems.\\ In order to understand the physical meaning of the 
	classical trajectories $\tau(E)$, we consider $\hb$ at the border of the sector 
	$\mathbb{C}^0$. At 
	$\arg 
	\hb=\pi/4$ we can factorize the imaginary unit $i$ and consider  the real cubic  
	oscillator
		\be
	H_r(\hb)=p^2+V^r(x),\quad p=-i\hb\,(d\!/dx)\quad\,V^r(x)=x^3-x,\,\quad\hb>0
	\label{Hr}
	\ee 
	for 
	an energy value $E\in A_0=(-c,\,c\,)$, with $c=2/(3\sqrt 3)$. Here we are not
	concerned with the non 
	completeness of the problem at $-\infty$. We get the classical Hamiltonian $H_r(p,x)$
	by substituting  in  $H_r(\hb)$ classical momentum $p$ to the operator $-i\hb\,(d/dx)\,.$
	The  union $\tau(E)$ of the 
	classical trajectories at energy $E\in A_0$ consists of the  
	{oscillation range} $\rho(E)=[I_-(E),I_+(E)]$  and the  {escape line} 
	$$\eta=\eta(E)=(-\infty,I_0(E)],\,\quad\,I_0(E)<-1/\sqrt 3<I_-(E)<I_+(E).$$  
	The real potential makes clear the meaning of our definitions of 
    oscillatory range $\rho(E)$ and of escape line $\eta(E)$ previously 
    used with complex potential. 
	Notice that $\tau(E)$ is unstable at  $E=c$, where  $\rho(E)$ touches  
	$\eta(E)$. \\
	Going back to the Hamiltonians $H_\hb$ at the limit cases $\arg(\hb)=\pm i(\pi/4)^-$, we have 
	the critical values of the energy $$E^c\bigl(\arg(\hb)=\pm(\pi/4)^-\bigr)=\pm ic=\mp 
	E_0=E_n^\mp(0).$$ 
	Thus we expect the existence of an infinite set of crossings in the complex $\hb$ plane 
	with complex accumulation points $E^c(\theta)$ of the crossing energies for 
	$|\hb|\ra 0$ along the direction $\arg (\hb)=\theta\neq 0$.
	About the discussion on all the crossings in the $\mathbb{C}^0$ complex sector of the $\hb$ 
	variable, see \cite{GG}. We expect  the generalized crossing rules in terms of the four limits  
	$$E_{m^\pm}(\hb_{(n^-,n^+)}+\es) - E^\pm_{m^\pm}(\hb_{(n^-,n^+)}-\es)\ra 
	0,\,\,\,\textrm{as}\,\,\,\es\ra 0,\,\,\,m^\pm=n^-+n^++(1\pm1)/2,\,\,\,$$ at $\hb_{(n^-,n^+)}$, 
	where 
	$\hb_{(n,n)}=\hb_n$. Thus, we expect as a general rule, the instability of one of the nodes of 
	the state $\psi_{m^+}(\hb)$ and the 
	partition between  the  two states  $\psi^\pm_{n^\pm}(\hb)$ of the $n^-+n^+$ nodes.\\The 
	hypothesis of 
	minimality about the singularities  gives the following picture. The analytic function 
	$E_{m}(\hb)$ for large $|\hb|$ has  
	a sequence of singularities  in $\mathbb{C}^0$ ordered by the increasing values of
	$\Im\hb_{j,k}$: 
	\be
	\hb_{m,0},\,\hb_{m-1,0}\,,\hb_{m-1,1},\,\hb_{m-1,2},....,\hb_{1,m-2},\,\hb_{1,m-1},\,
	\hb_{0,m-1},\,\hb_{0,m}
	\label{Seq_h}
	\ee
	It is possible to   divide $\mathbb{C}^0$, for small $|\hb|$, by cuts going from $0$  to the 
	branch points (\ref{Seq_h})
	in a sequence of stripes   
	$$S_m^-,\,S_0^+,\,S^-_{m-1},\, S^+_{1},....,S^-_{1},\,S^+_{m-1},\,S^-_{0},\,S^+_{m}$$ 
	in which, for $\hb\ra0$, the level $E_m(\hb)$ 
	has the behaviors 
	$$E_m^-(\hb),\,E_0^+(\hb),\,E^-_{m-1}(\hb),\, 
	E^+_{1}(\hb),....,E^-_{1}(\hb),\,E^+_{m-1}(\hb),\,E^-_{0}(\hb),\,E^+_{m}(\hb)$$ 
	respectively. 
\\\\
	
	\bigskip
	
	\textbf{Aknowlodgements.} It is a pleasure to thanks Professor Andr\'e Martinez for many 
	suggestions
	and for giving us the reference \cite {GI}.
	\vfill\break

	\end {document}